\documentclass[aps,preprint,superscriptaddress,preprintnumbers,nofootinbib]{revtex4}
\usepackage{amsmath,amssymb,graphicx,bm}
\usepackage{enumerate}
\usepackage{subfig}
\usepackage{hyperref}
\usepackage{color}

\newcommand{\met}{{E\!\!\!\!/_{\rm T}}}

\newcommand{\beq}{\begin {equation}}  
\newcommand{\eeq}{\end   {equation}} 
\newcommand{\bea}{\begin {eqnarray}} 
\newcommand{\eea}{\end   {eqnarray}}  
\newcommand{\baa}{\begin {array}   } 
\newcommand{\eaa}{\end   {array}   }     
\newcommand{\bit}{\begin {itemize} }
\newcommand{\eit}{\end   {itemize} }
\newcommand{\be }{\begin {equation}} 
\newcommand{\ee }{\end   {equation}}

\newcommand{\tbox}[1]{\mbox{\tiny #1}}

\begin{document}

\preprint{UTTG-11-15}
\preprint{MI-TH-1522}
\preprint{CETUP2015-007}

\title{G221 Interpretations of the Diboson and $Wh$ Excesses}

\author{Yu Gao}
\author{Tathagata Ghosh}
\affiliation{Mitchell Institute for Fundamental Physics and Astronomy, Department of Physics and Astronomy, Texas A\&M University, College Station, TX 77843-4242, USA}
\author{Kuver Sinha}
\affiliation{Department of Physics, Syracuse University, Syracuse, NY 13244, USA}
\author{Jiang-Hao Yu}
\affiliation{Theory Group, Department of Physics and Texas Cosmology Center,
\\The University of Texas at Austin,  Austin, TX 78712, USA}

\date{\today}

\begin{abstract}
Based on an $SU(2) \times SU(2) \times U(1)$ effective theory framework (aka $G221$ models),  we investigate a leptophobic $SU(2)_L \times SU(2)_R \times U(1)_{B-L}$ model, in which the right-handed $W^\prime$ boson has the mass of around 2 TeV, and predominantly couples to the standard model quarks and the gauge-Higgs sector. This model could explain the resonant excesses near 2 TeV reported by the ATLAS collaboration in the $WZ$ production decaying into hadronic final states, and by the CMS collaboration in the $Wh$ channel decaying into $b\bar{b}\ell\nu$ and dijet final state. After imposing the constraints from the electroweak precision and current LHC data, we find that to explain the three excesses in $WZ$, $Wh$ and dijet channels, the $SU(2)_R$ coupling strength $g_R$ favors the range of $0.47 \sim 0.68$. In this model, given a benchmark 2 TeV $W'$ mass, the $Z'$ mass is predicted to be around $2.9$ TeV if the doublet Higgs (LPD) is used to break the G221 symmetry, consistent with the 2.9 TeV $e^+e^-$ event recently observed at CMS. A $3 \sim 5$ TeV mass is typically predicted for the triplet Higgs (LPT) symmetry breaking scenario, can also be consistent with a 2.9 TeV dilepton signal. These signatures can be further explored by the LHC Run-2 data. 
\end{abstract}

\maketitle

\section{Introduction}
\label{sec:intro}

The ATLAS collaboration has recently reported excesses in searches for massive resonances decaying into a pair of weak gauge bosons~\cite{Aad:2015owa}. The anomalies have been observed in all hadronic  final states in the $WZ$, $WW$, and $ZZ$ channels at around 2 TeV invariant mass of the boson pair. The analysis has been done with 20.3 fb$^{-1}$ of data at 8 TeV, with local significances of 3.4 $\sigma$, 2.6 $\sigma$, and 2.9 $\sigma$ in the $WZ$, $WW$, and $ZZ$ channels, respectively. Several groups~\cite{Fukano:2015hga} have studied this excess. Similar moderate diboson excesses have also been reported from the CMS~\cite{Khachatryan:2014hpa,Khachatryan:2014gha} experiment. Intriguingly, the CMS experiment reported around 2$\sigma$ excesses slightly below 2 TeV in the dijet resonance channel~\cite{Khachatryan:2015sja} and $e\nu b\bar{b}$~\cite{CMS:2015gla} channel which may arise from a $W'\rightarrow Wh$ process.

A natural question to ask is whether a single resonance whose peak is around 2 TeV and width less than 100 GeV can nicely fit all the excesses. The tagging selections used in the analysis do not give a completely clear answer - around 20$\%$ of the events are shared among the three channels~\cite{Aad:2015owa}, leading to the possibility of cross contamination. While a single resonance is definitely the simplest option, a more realistic possibility is that several resonances are present at the 2 TeV mass scale, where new physics presumably kicks in. The most natural scenario is then that these resonances are associated with the spontaneous breaking of extra gauge groups at that scale. Scalars in the extra sectors, for example, would need significant mixing with the Standard Model Higgs to be reproduced at the LHC and give the observed excesses. The other option is that the resonances are gauge bosons of the new gauge groups, which acquire mass through a Higgs mechanism in the extra sector. This is the avenue we pursue in this paper. 

There are several immediate caveats when one considers this possibility. Firstly, extra gauge bosons will decay to the diboson channels through their mixing with the SM $W$ and $Z$. Such mixing is constrained by electroweak (EW) precision tests, necessitating the balance between obtaining the correct cross-section to fit the excesses and accommodating EW constraints.  The second caveat is that the SM fermions can be charged under the extra gauge group and let the exotic gauge bosons decay into SM fermionic states. One then has to be careful about dilepton and dijet constraints for such a resonance, with the possibility that the former is evaded by working in the context of a leptophobic model. Thirdly, the excess in the $ZZ$ channel cannot be accounted for only with exotic gauge bosons. This makes such scenarios falsifiable in the near future; the persistence of the excess in the $ZZ$ channel would indicate extra physics at the 2 TeV scale, apart from the exotic gauge bosons considered here. 

The purpose of this paper is to investigate exotic gauge bosons $W'$ as a candidate 
for the 2 TeV resonance in the light of the caveats mentioned above. 
In extended gauge group models, usually both the $W'$ boson and the $Z'$ bosons exists. 
We would like to focus on the low energy effective theory of extended gauge group models 
, in which all the heavy particles other than the $W^\prime$ and $Z^\prime$ bosons decouple. 
This has been studied in the $SU(2)\times SU(2)\times U(1)$ framework, as the so-called G221 models~\cite{Hsieh:2010zr,Cao:2012ng}.
The $G221$ models are the minimal extension of the SM gauge group to incorporate both the $W^\prime$ 
and $Z^\prime$ bosons.  
Various models have been considered under this broad umbrella:  
left-right (LR)~\cite{Mohapatra:1974gc, Mohapatra:1974hk, Mohapatra:1980yp},
lepto-phobic (LP), hadro-phobic (HP), fermio-phobic (FP)~\cite{Hsieh:2010zr, Cao:2012ng,  Chivukula:2006cg, Barger:1980ix, Barger:1980ti}, un-unified (UU)~\cite{Georgi:1989ic, Georgi:1989xz}, and non-universal (NU)~\cite{Li:1981nk, Muller:1996dj, Malkawi:1996fs, He:2002ha, Berger:2011xk}.

As an explicit model, we will focus on the leptophobic (LP) G221 model with two stage symmetry breaking. 
In the first stage breaking, a doublet Higgs (LPD) or a triplet Higgs (LPT) could be introduced. 
In this model, the $W^\prime$ boson couplings to the SM leptons are highly suppressed. 
Therefore, this leptophoic model could escape the tight constraints from lepton plus missing energy searches.
At the same time, the $W^\prime$ boson couplings to the SM quarks and gauge bosons are similar to the typical left-right model. 
Therefore, the $W'$ can be produced at the LHC with potentially large production rate, and mainly decay to the  
dijet, $t\bar{b}$, $WZ$ and $Wh$ final states, instead of the $\ell \nu$ final states. 
We will explain the resonant excesses near 2 TeV reported by the ATLAS collaboration in the $WZ$ production decaying into hadronic final states, and
by the CMS collaboration in the $Wh$ channel decaying into $b\bar{b}\ell\nu$ and dijet final state. 
Given the $W'$ mass at 2 TeV and expected signal rate on the $WZ$ final state, the model parameters are fixed. 
Therefore, we predict the $Z'$ mass and couplings to the SM particles. 
For the LPD model, the $Z'$ mass is predicted to favor $2 \sim 3 $ TeV, while $3 \sim 5$ TeV for the LPT model. 
Unlike to the $W'$ boson which is totally leptophobic, the $Z'$ will couple to the SM leptons due to the extra $U(1)$ charge. The CMS experiment has recently reported a 2.9 $e^+e^-$ event~\cite{bib:cms2.9dielectron} that can be well explained by the $Z'$ resonance in both LPD and LPT models. Even if only previous no-signal data in dilepton searches are considered as a constraint $m_{Z'} < 2.7 \sim 2.8$ TeV, the LPT model can still be consist with its heavier $Z'$ mass prediction.
We also include the electroweak precision constraints in the parameter space. 
Although some parameter region of the LPD model might be highly constrainted due to the dilepton final states, the LPT model could satisfy all the constraints and explain the $WZ$, $Wh$ and dijet excesses.

The rest of the paper is structured as follows. In Section \ref{sec:model}, we describe the model in detail. In Section \ref{sec:constraint}, we describe the constraints on our model coming from electroweak precision tests. In Section \ref{sec:explain}, we describe our main results and predictions. We end with our conclusions.


\section{The $SU(2) \times SU(2) \times U(1)$ Model}
\label{sec:model}

As mentioned in the introduction, we will be explicitly working in the context of the $G221$ models~\cite{Hsieh:2010zr,Cao:2012ng}, which we now briefly review. 
The $G221$ models are the minimal extension of the SM gauge group to incorporate both the $W^\prime$ 
and $Z^\prime$ bosons. 
This model can be treated as the low energy effective theory of extended gauge group models 
with all the heavy particles other than the $W^\prime$ and $Z^\prime$ bosons decouple. 
The gauge structure is $SU(2)\times SU(2)\times U(1)$.
There are two kinds of breaking patterns: the
$SU(2)\otimes U(1)$ breaking down to $U(1)_{Y}$ (breaking pattern I, where the $W^\prime$ mass
is smaller than the $Z^\prime$ mass), and 
the $SU(2)\otimes SU(2)$ breaking down to $SU(2)_{L}$ (breaking pattern II, where the
$W^\prime$ and $Z^\prime$ bosons have the same mass).
In the breaking pattern I, the model structure is the left-right symmetry $SU(2)_L \times SU(2)_R \times U(1)_X$ with different
charge assignments in fermion sector,
while in the breaking pattern II, the model includes two left-handed $SU(2)$ with $SU(2)_{L1} \times SU(2)_{L2} \times U(1)_Y$ gauge structure and  different 
charge assignments in fermion sector.
We will mainly be interested in the lepto-phobic (LP) model.
In this model, the following symmetry breaking pattern (breaking pattern I) is applied with gauge structure $SU(2)_L \times SU(2)_R \times U(1)_X$. 
In the first stage, 
the  breaking  $SU(2)_R \times U(1)_{X} \rightarrow U(1)_{Y}$ occurs at the $\sim$ 2 TeV scale, while the second stage of 
symmetry breaking $SU(2)_{L} \times U(1)_Y \rightarrow U(1)_{em}$ takes place at the EW scale.

The gauge couplings for $SU(2)_L$, $SU(2)_R$, and $U(1)_{X}$ are denoted by $g_L$, $g_R$ and $g_X$, respectively. 
In the above notation, the gauge couplings are given by
\be
g_L = \frac{e}{\sin\theta}, \,\,\,\, g_R = \frac{e}{\cos\theta \sin{\phi}}, \,\,\,\, g_X = \frac{e}{\cos\theta \cos{\phi}} \,\,.
\ee
where the couplings are correlated by the SM weak mixing angle $\theta$ a new mixing angle $\phi$. 
In this model, the SM left-handed fermion doublets are charged under the $SU(2)_L$, 
the right-handed quark doublet are charged under the  $SU(2)_R$. 
We identify the $U(1)_{X}$ as the $U(1)_{B-L}$ gauge symmetry in the following.
The charge assignments of the SM fermions are shown in Table~\ref{tb:models}.

\begin{table}[h]
\begin{center}
\caption{
The charge assignments of the SM fermions under
the leptophobic $G221$ model.
}
\label{tb:models}
\vspace{0.125in}
\begin{tabular}{|c|c|c|c|}
\hline Model & $SU(2)_L$ & $SU(2)_R$ & $U(1)_{B-L}$ \\
\hline
\hline
Lepto-phobic &
$\begin{pmatrix} u_L \\ d_L \end{pmatrix}, \begin{pmatrix} \nu_L \\ e_L \end{pmatrix}$ &
$\begin{pmatrix} u_R \\ d_R \end{pmatrix}$ &
$\begin{matrix} \tfrac{1}{6}\ \mbox{for quarks,} \\ Y_{\tbox{SM}}\ \mbox{for leptons.} \end{matrix}$
\\
\hline
\end{tabular}
\end{center}
\end{table}

At the TeV scale, the $SU(2)_{R} \times U(1)_{B-L} \rightarrow U(1)_{Y}$ breaking can be induced by a scalar doublet $\Phi\sim(1,2)_{1/2}$ (LPD) or a scalar triplet $(1,3)_{1}$ (LPT)~\footnote{the quantum number assignment is under $(SU(2)_L, SU(2)_R)_{U(1)_{B-L}}$} with a vacuum expectation value
(VEV) $u$. 
Another bi-doublet scalar is introduced for the subsequent $SU(2)_{L} \times U(1)_{Y} \rightarrow U(1)_{Q}$ at the EW scale. 
This is denoted by $H \sim (2,\bar{2})_{0}$ with two VEVs $v_1$ and $v_2$. 
We will prefer to change variables and work with a single VEV  $v = \sqrt{v_1^2 + v_2^2}$ and a mixing angle $\beta = \arctan(v_1/v_2)$. 
We define a quantity $x$, which is the ratio of the VEVs
\be
x \,\, = \,\, \frac{u^2}{v^2} \,\,,
\ee
with $x \, \gg \, 1$.
Usually the physical observables are not sensitive to the parameter $\beta$ 
as it contributes to physical observables only at the order of $1/x$. 
So in the following discussion, we will fix $\sin2\beta$ to be one to maximize the 
$W'$ couplings to the gauge bosons and the Higgs boson.

The gauge bosons of the $G221$ model are denoted by
\begin{align}
SU(2)_L&: W_{1,\mu}^{\pm}, W_{1,\mu}^{3},\nonumber\\
SU(2)_R&: W_{2,\mu}^{\pm}, W_{2,\mu}^{3},\nonumber\\
U(1)_{B-L}&: X_{\mu}.
\end{align}
After symmetry breaking, both $W^\prime$ and $Z^\prime$ bosons obtain masses
and mix with the SM gauge bosons. 
To order $1/x$ the eigenstates of the charged gauge bosons are
\begin{eqnarray}
W_\mu^\pm &=& {W_1^\pm}_\mu +\frac{\sin\phi \sin2\beta}{x
\tan\theta}{W_2^\pm}_\mu \, ,\\
{W^\prime}_\mu^\pm &=&  -\frac{\sin\phi \sin2\beta}{x \tan\theta}
{W_1^{\pm}}_{\mu}+{W_2^{\pm}}_{\mu}  \, .
\end{eqnarray}
While for the neutral gauge bosons
\begin{eqnarray}
Z_\mu &=& {W_Z^3}_\mu +\frac{\sin \phi \cos^3 \phi}{x\sin \theta}
                           {W_H^3}_\mu\, ,\\
Z_\mu^{\prime} &=&  -\frac{\sin\phi\cos^3\phi}{x\sin\theta}
{W_Z^3}_\mu + {W_H^3}_\mu \, ,
\end{eqnarray}
where $W_H^3$ and $W_Z^3$ are defined as
\begin{eqnarray}
{W_H^3}_{\mu} &=& \cos\phi   {W_2^3}_{\mu} - \sin\phi X_{\mu}\,,\\
{W_Z^3}_{\mu} &=& \cos\theta {W_1^3}_{\mu} -\sin\theta (\sin\phi {W_2^3}_{\mu} + \cos\phi X_{\mu})\,,\\
{A    }_{\mu} &=& \sin\theta {W_1^3}_{\mu} +\cos\theta (\sin\phi {W^3_2}_{\mu} + \cos\phi X_{\mu}).
\end{eqnarray}
Correspondingly, 
the masses of the $W^{\prime}$ and $Z^{\prime}$ are given by
\bea
M_{{W^{\prime}}^{\pm}}^{2} = \frac{e^{2}v^{2}}{4\cos^{2}\theta \sin^{2}{\phi}}\left(x+1\right)\,,
\quad
M_{Z^{\prime}}^{2}  = \frac{e^{2}v^{2}}{4\cos^{2}\theta\sin^{2}{\phi}\cos^{2}{\phi}}\left(x+\cos^{4}{\phi}\right)\,,
\label{mzp_bp1}
\eea
for the LPD model,
and
\bea
M_{{W^{\prime}}^{\pm}}^{2} = \frac{e^{2}v^{2}}{4\cos^{2}\theta \sin^{2}{\phi}}\left(2x+1\right)\,,
\quad
M_{Z^{\prime}}^{2}  = \frac{e^{2}v^{2}}{4\cos^{2}\theta\sin^{2}{\phi}\cos^{2}{\phi}}\left(4x+\cos^{4}{\phi}\right)\,,
\label{mzp_bp1t}
\eea
for the LPT model.

For the LPD, the relevant Feynman rules on the fermion couplings are written as
\begin{equation}
W^{\prime\pm} \overline{f}f':\quad \frac{e}{\sqrt{2} \sin\theta} \left(f_{W'L} P_L + f_{W'R} P_R	\right),
\end{equation}	
with
\begin{eqnarray}
f_{W'L} = -\frac{\sin\phi \sin(2\beta)}{x \tan\theta}, \quad \quad f_{W'R} = \frac{\tan\theta}{\sin\phi},	
\end{eqnarray}
and 
\begin{equation}
Z^{\prime}\overline{f}f: \quad \frac{e}{\sin\theta\cos\theta} \left(f_{Z'L} P_L + f_{Z'R} P_R	\right),
\end{equation}	
with
\begin{eqnarray}
f_{Z'L} &=& (T^3-Q) \sin\theta\tan\phi-(T^3-Q\sin^2\theta)\frac{\sin\phi \cos^3\phi}{x \sin\theta} \\
f_{Z'R} &=& (T^3-Q\sin^2\phi) \frac{\sin\theta}{\sin\phi\cos\phi} +Q\frac{\sin\theta\sin\phi\cos^3\phi}{x}.
\end{eqnarray}

For the LPD, the gauge boson self-couplings are given as follows,
with all momenta out-going.  The three-point couplings take the form:
\begin{equation}
	V_1^{\mu}(k_1) V_2^{\nu}(k_2) V_3^{\rho}(k_3): \ \ 
	- i f_{V_1V_2V_3} \left[ g^{\mu\nu}(k_1-k_2)^{\rho}
	+ g^{\nu\rho}(k_2-k_3)^{\mu}
	+ g^{\rho\mu}(k_3-k_1)^{\nu} \right],
\end{equation}
where the coupling strength $f_{V_1V_2V_3}$ for the $WWZ'$ and $W'WZ$ are
\begin{eqnarray}
 f_{WWZ'} = \frac{e \sin\phi \cos^3\phi \cot\theta}{x \sin\theta},\quad
 f_{W'WZ} = \frac{e \sin\phi \sin(2\beta)}{x \sin^2\theta}.
\end{eqnarray}

Similarly, the $HWW'$ and $HZZ'$ couplings in the LPD are 
\begin{eqnarray}
 HWW': g^{\mu\nu}\frac{e^2 v}{2\sin^2\theta}  f_{HWW'}, \quad
  HZZ':  g^{\mu\nu}\frac{e^2 v}{2\sin^2\theta\cos^2\theta}  f_{HZZ'},
\end{eqnarray}
with the coupling strengths are
\begin{eqnarray}
    f_{HWW'} &=& -\frac{\sin(2\beta)\tan\theta}{\sin\phi}+\frac{\sin(2\beta)(\tan\theta-\cot\theta\sin^2\phi)}{x \sin\phi},\\
	f_{HZZ'} &=& -\frac{\sin\theta}{\tan\phi} +\frac{\cos^3\phi( \sin^2\theta \cos^2\phi -\sin^2\phi )}{x \sin\theta \sin\phi}.
\end{eqnarray}
For the LPT Feynman rules, the only change on the couplings to the fermion, gauge and Higgs bosons  is that  replacing $x$ to $2x$ for the  $W'$ couplings,
and replacing $x$ to $4x$ for the $Z'$ couplings.

\begin{figure}
	\includegraphics[width=0.43\textwidth]{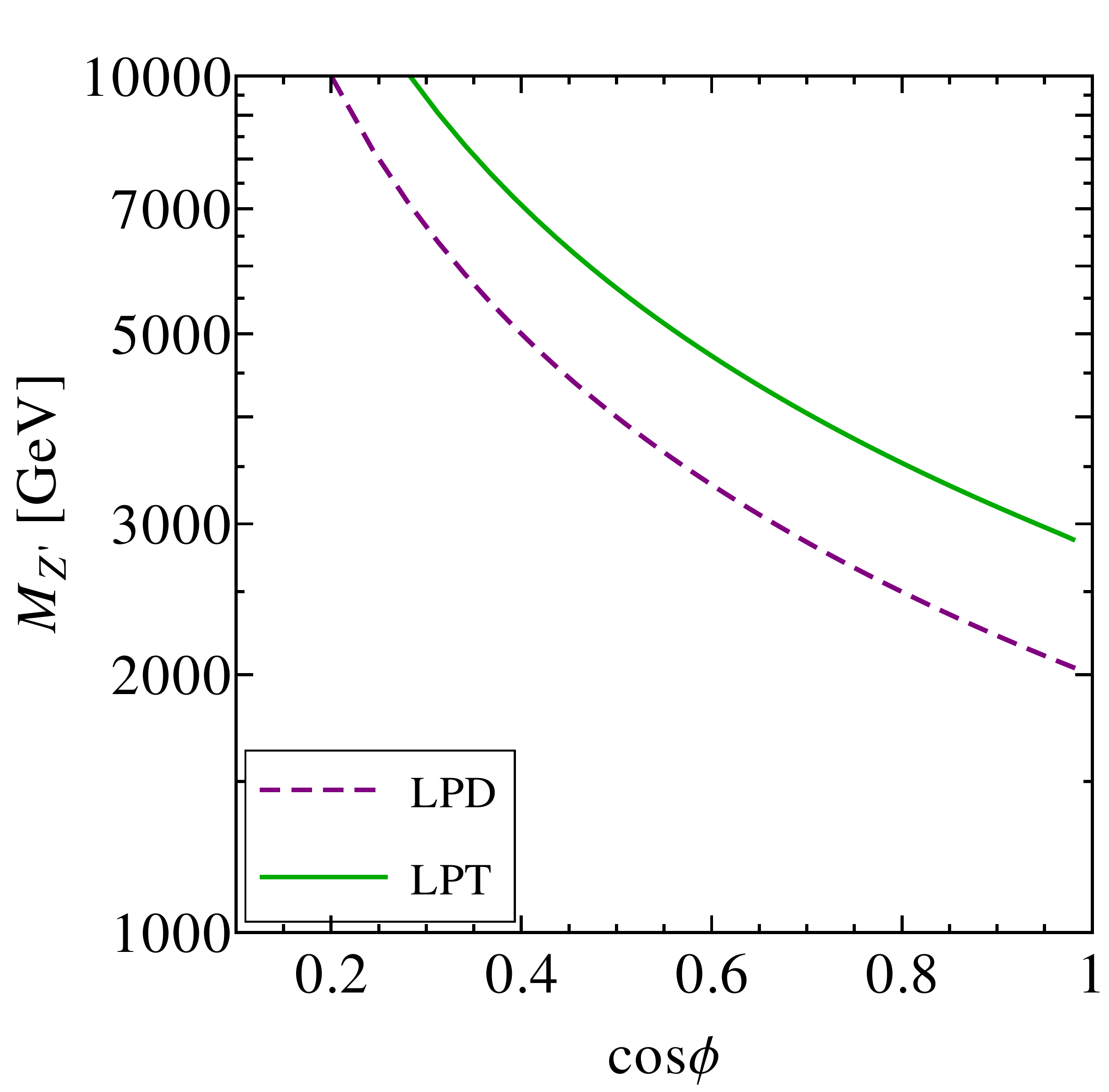}
	\caption{Given the $W'$ mass at 2 TeV, the  $Z^{\prime}$ masses in the lepto-phobic doublet (LPD) model and the lepto-phobic triplet (LPT) as a function of the mixing angle $\cos{\phi}$.}
\label{zprime_mass}
\end{figure}

\begin{figure}
	\includegraphics[width=0.43\textwidth]{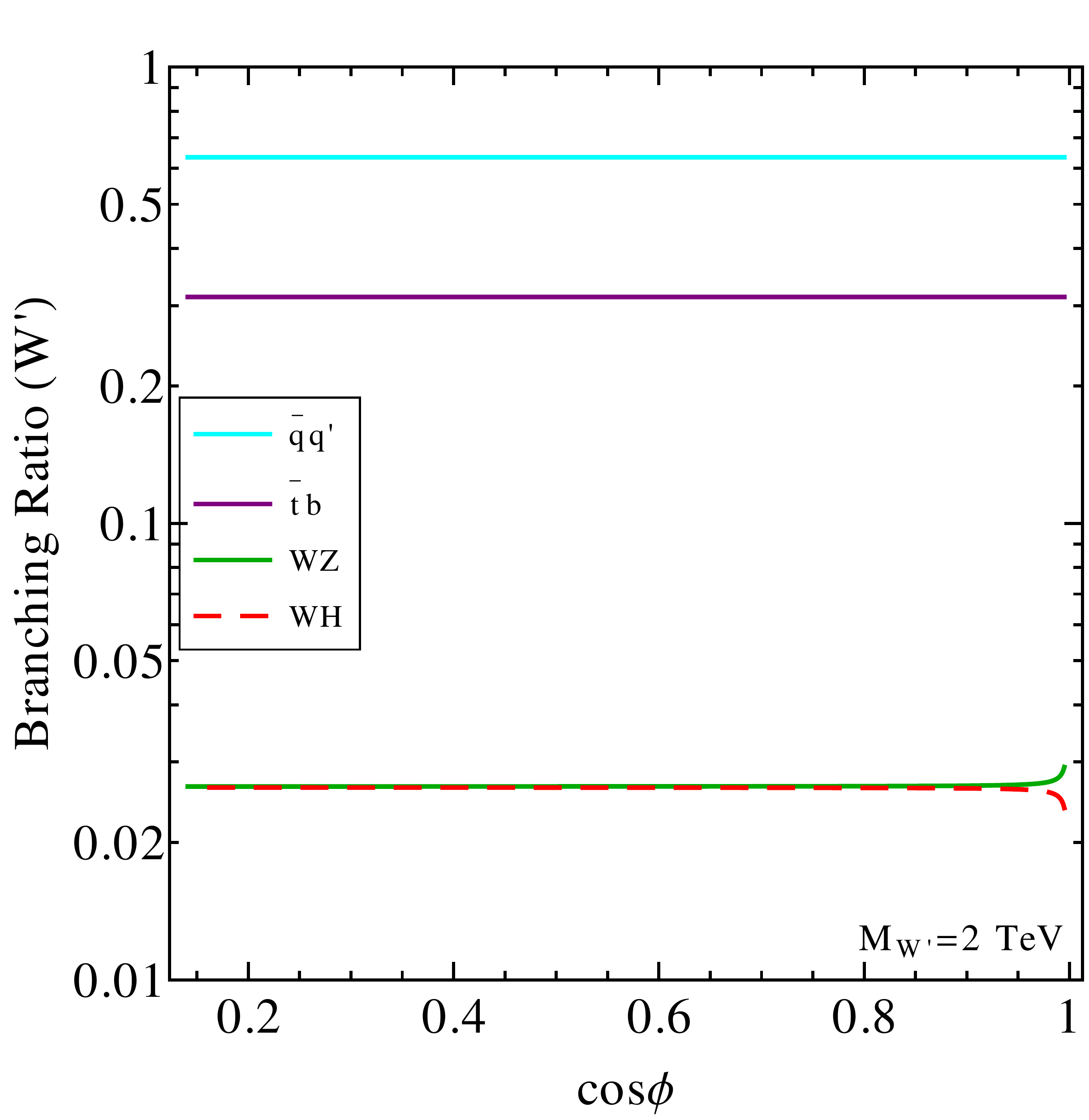}
	\includegraphics[width=0.45\textwidth]{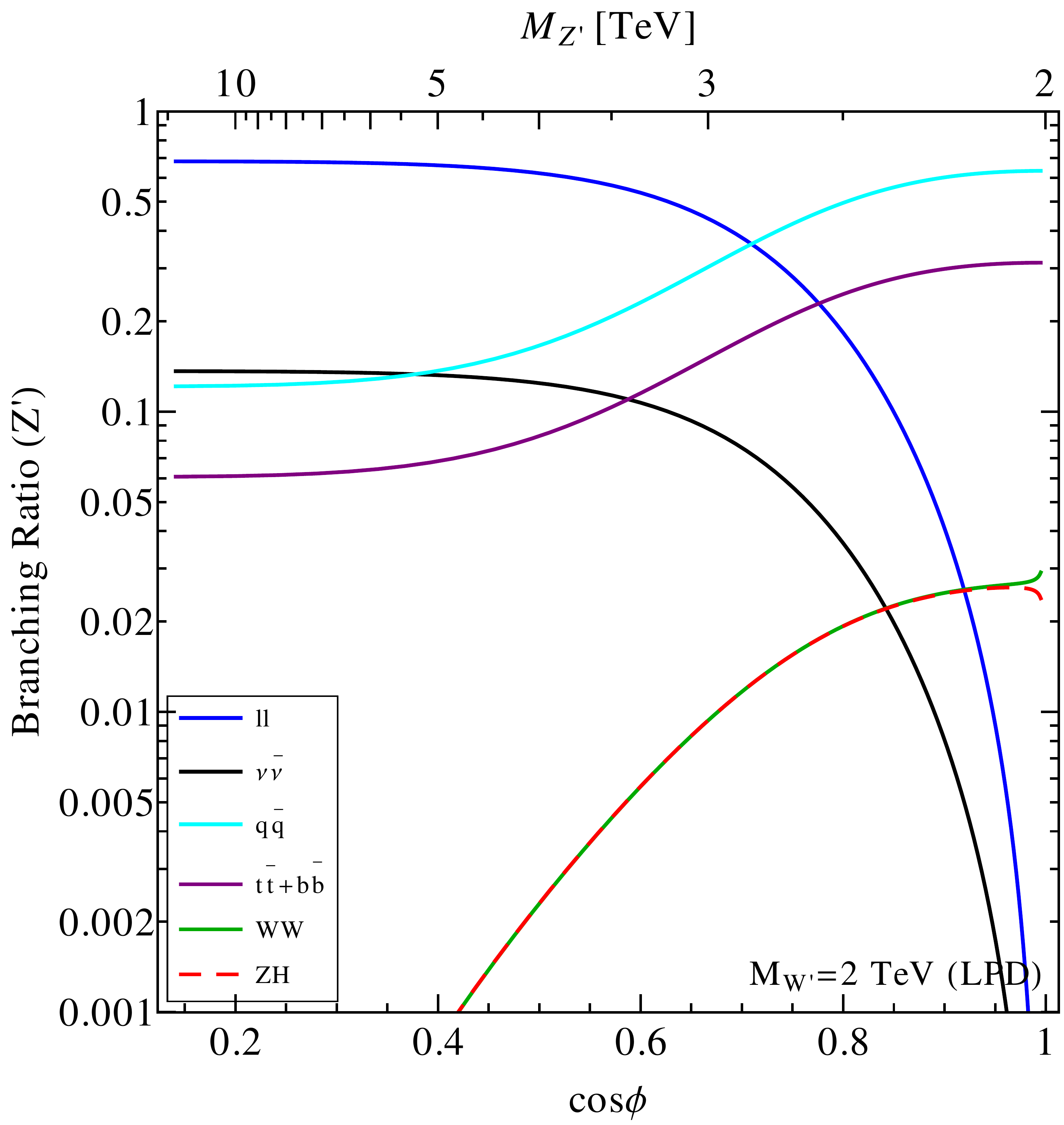}
	\caption{The branchings of $W^{\prime}$ (left column) and $Z^{\prime}$ (right column) to various SM states in the lepto-phobic doublet (LPD) $G221$ model at the point $M_{W^{\prime}} \, = \, 2$ TeV, as a function of the mixing angle $\cos{\phi}$.}
\label{branchings_LPD}
\end{figure}

According to the above Equations \ref{mzp_bp1} and \ref{mzp_bp1t}, the $W'$ mass and the $Z'$ mass are strongly correlated through the mixing angle $\cos{\phi}$. 
Given the $W'$ mass and the mixing angle $\cos{\phi}$, the $Z'$ mass is fully determined.
The Figure \ref{zprime_mass} shows 
when the $W'$ mass is at 2 TeV the  $Z^{\prime}$ masses in LPD and LPT models as a function of the mixing angle $\cos{\phi}$.
As a benchmark point, we will pick up the mixing angle  $\cos{\phi} = 0.8$ with $M_{W^{\prime}} \, = \, 2$ TeV.
Using the above Feynman rules, one can calculate the decay width and branching ratios of $W^{\prime}$ and $Z^{\prime}$ to various SM states. The details are shown in the Appendix. For future reference, we display below the branchings for $W^{\prime}$ and $Z^{\prime}$ at the point $M_{W^{\prime}} \, = \, 2$ TeV and $M_{Z^{\prime}} \, = \, 2.9$ TeV, which corresponds to the benchmark point with $\cos{\phi} = 0.8$, in the LPD $G221$ model.
From the Figure \ref{branchings_LPD}, we also see that the branching ratio $\textrm{Br}(W' \to WZ)$ is almost equal to the branching ratio $\textrm{Br}(W' \to Wh)$. 
This is because when the $W'$ is heavy, the decay product $W$ and $Z$ are highly boosted with the longitudinal polarization 
$\epsilon^\mu_L(k) \sim k^\mu$. According to the equivalence theorem, we know $\sigma(W' \to WZ) \sim \sigma(W' \to Wh)$.
Similarly we see that $\sigma(Z' \to WW) \sim \sigma(Z' \to Zh)$.

\section{Electroweak Precision Constraints}
\label{sec:constraint}

In this Section, we describe the constraints coming from EW precision tests (EWPTs) \cite{Abele:2001js, Erler:1999ug}.

In \cite{Hsieh:2010zr, Cao:2012ng}, a global-fit analysis of 37 EWPTs was performed to derive the allowed model parameter space in the LP $G(221G221)$ model \footnote{Since there is tree-level mixing between the extra gauge bosons and the SM gauge bosons, all the EWPT data cannot be described by the conventional oblique parameters $(S,T,U)$. A global fit is thus performed.}. From Eq. \ref{mzp_bp1}, it is clear that $M_{W^{\prime}}$ and $M_{Z^{\prime}}$ are not independent parameters. Therefore, $M_{W^{\prime}}$ was chosen as the input mass. The other independent parameters are the gauge mixing angle $\phi$ and the mixing angle $\beta$. Since the parameter scan is not very sensitive to the angle $\beta$, which becomes important only at $\mathcal{O}(1/x)$, it can be ignored. Thus, the scans will be presented in the $(M_{W^{\prime}}, c_{\phi})$ plane or the $(M_{W^{\prime}}, M_{Z^{\prime}})$ plane.

\begin{figure}
	\includegraphics[width=0.41\textwidth]{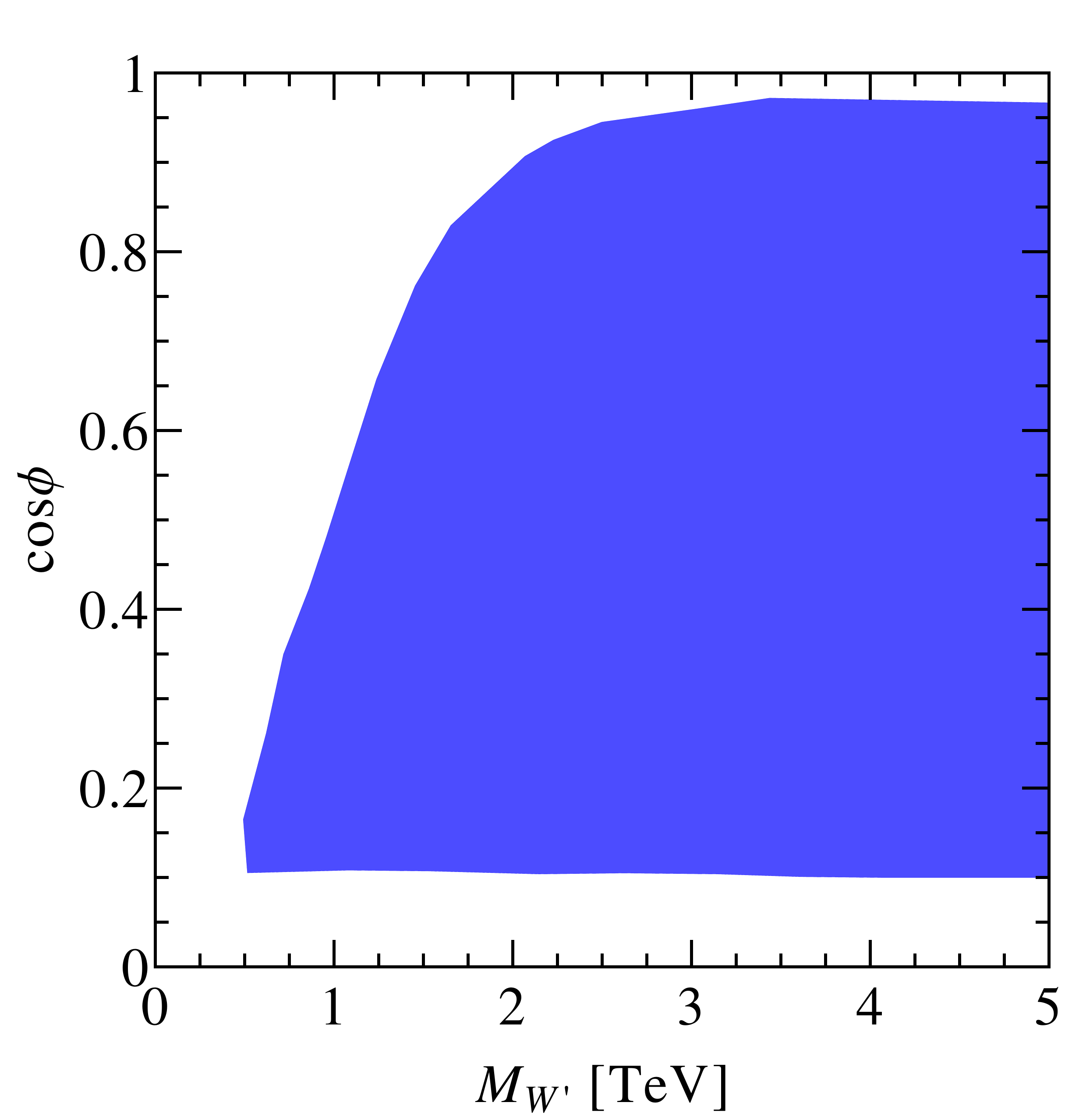}
	\includegraphics[width=0.4\textwidth]{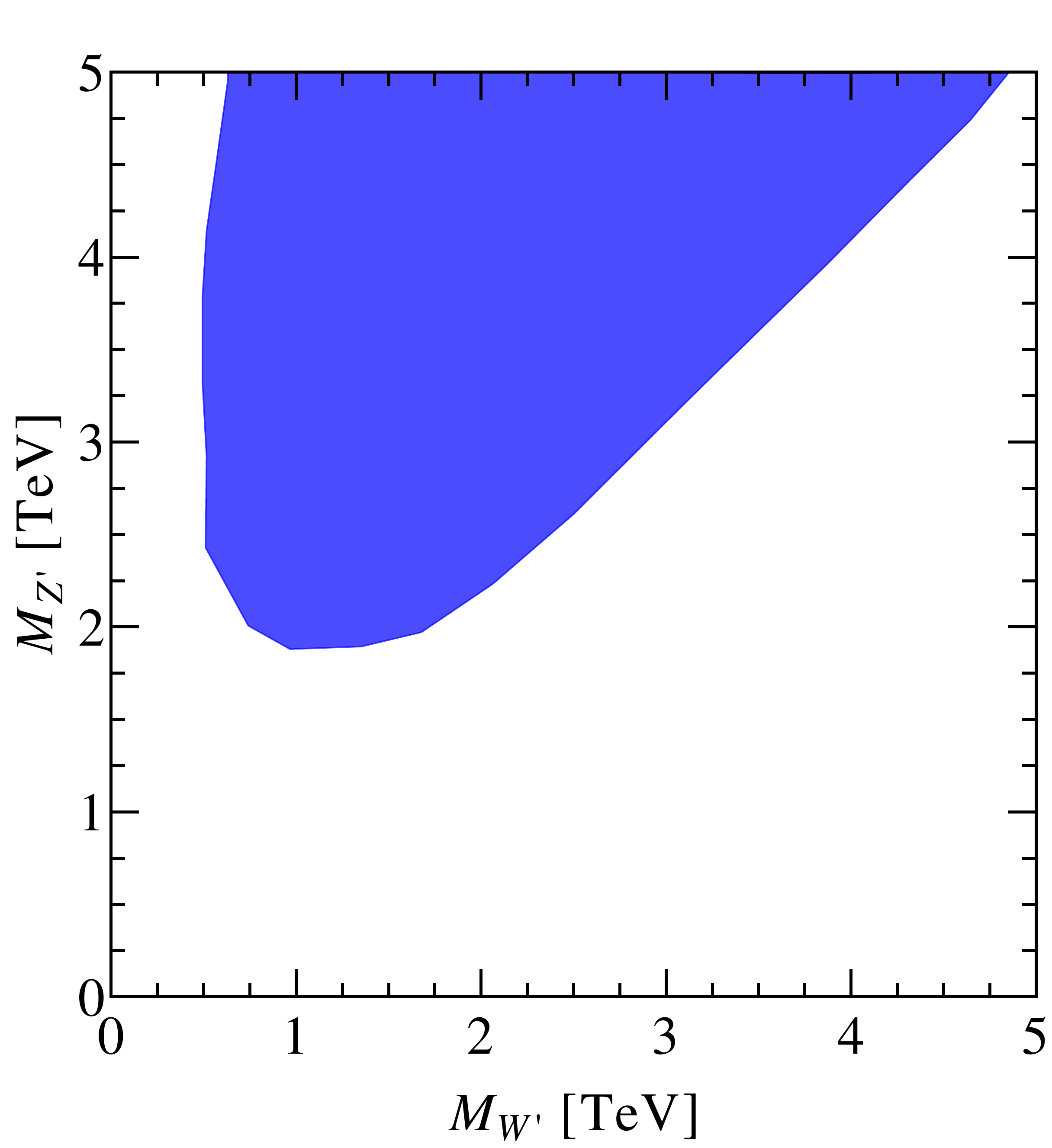}
	\caption{Allowed parameter space (blue colored region) of the lepto-phobic doublet (LPD) $G221$ model at $95\%$ CL  in the $\cos{\phi} - M_{W^\prime}$ and $M_{Z^\prime}-M_{W^\prime}$ planes after including EWPT constraints.}
\label{parameter_constraints_LPD}
\end{figure}

\begin{figure}
	\includegraphics[width=0.41\textwidth]{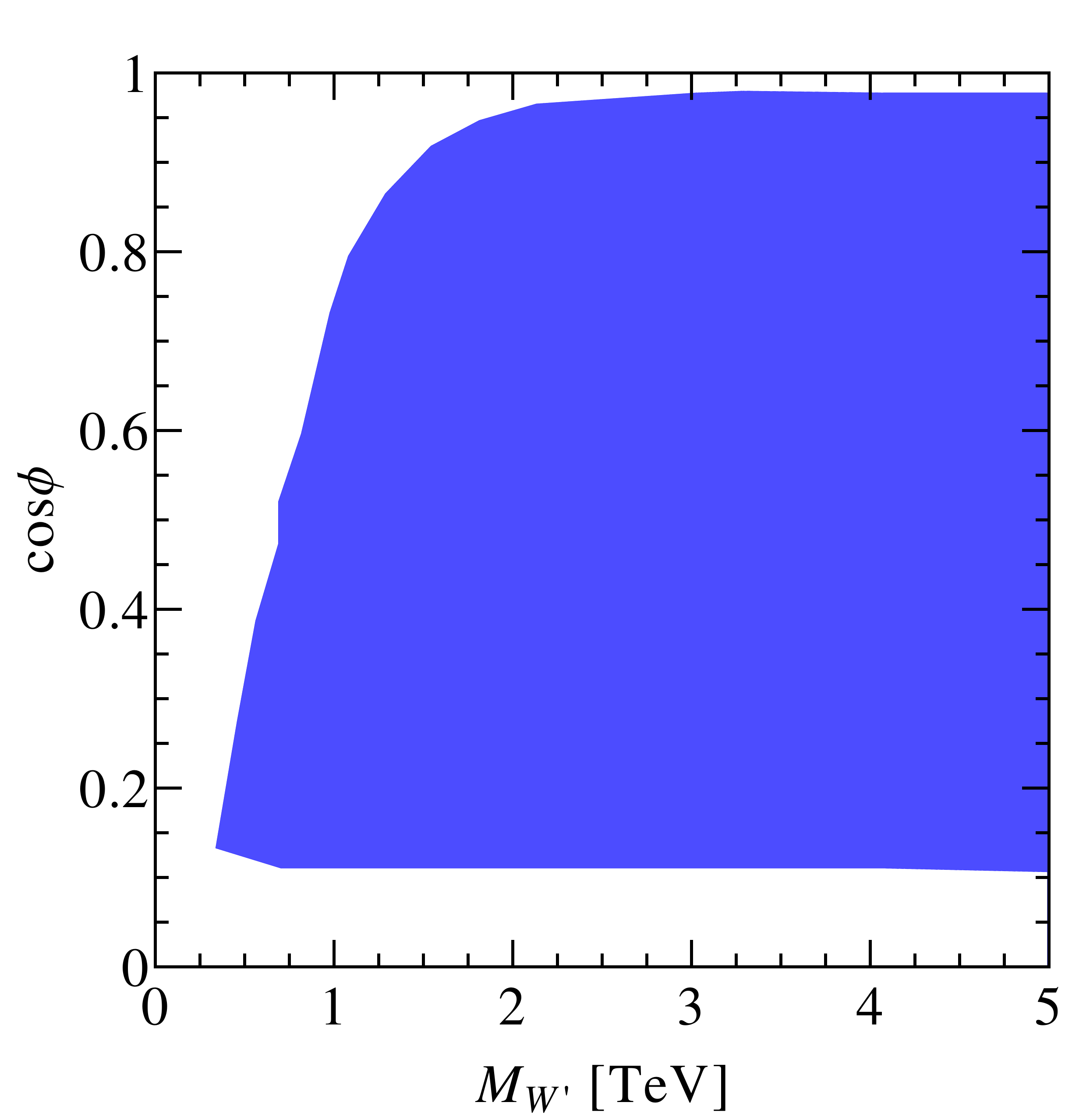}
	\includegraphics[width=0.4\textwidth]{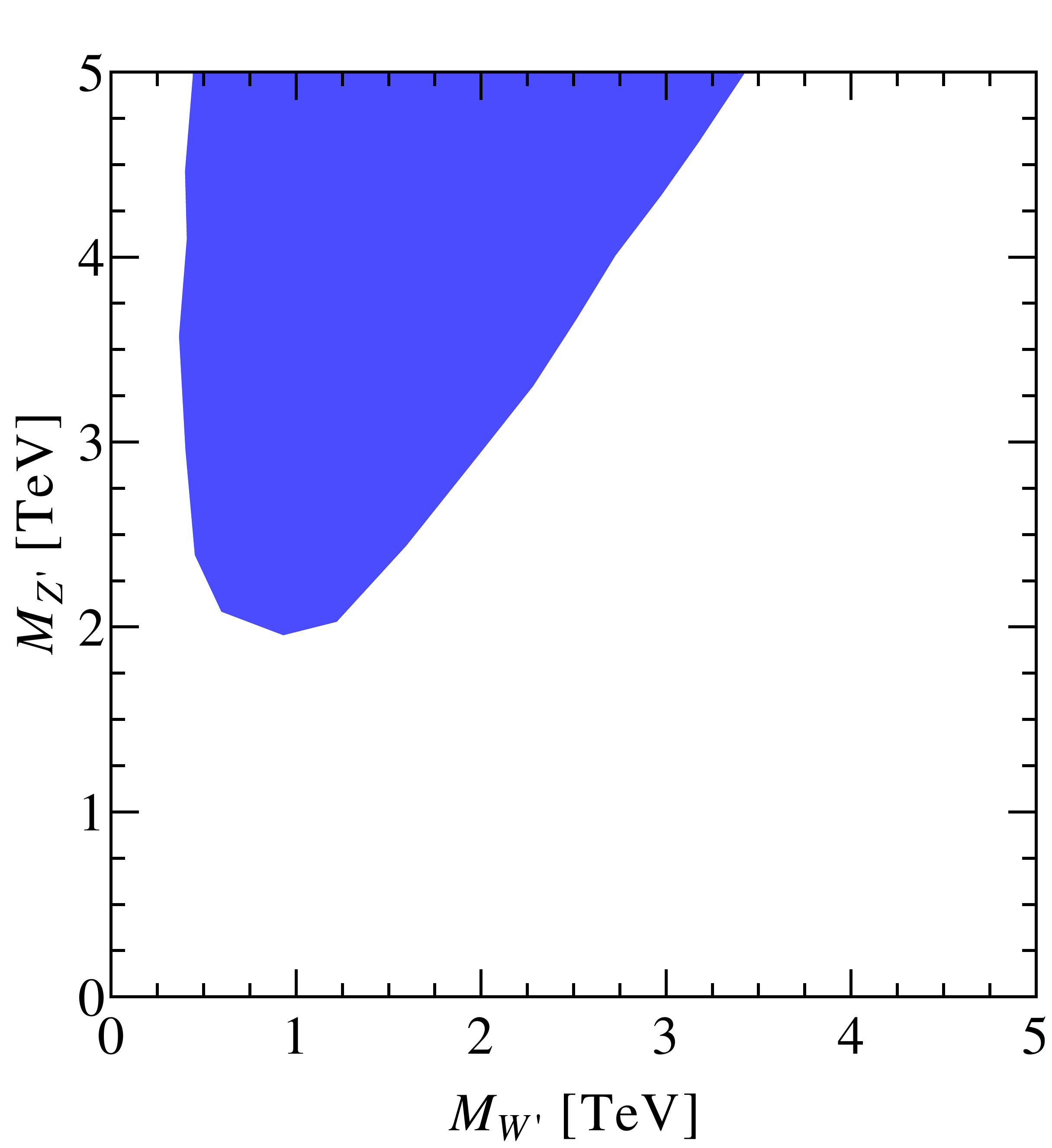}
	\caption{Allowed parameter space (blue colored region) of the lepto-phobic triplet (LPT) $G221$ model at $95\%$ CL  in the $\cos{\phi} - M_{W^\prime}$ and $M_{Z^\prime}-M_{W^\prime}$ planes after including EWPT constraints.}
\label{parameter_constraints_LPT}
\end{figure}

In Fig. \ref{parameter_constraints_LPD} and Fig. \ref{parameter_constraints_LPT}, we show the allowed parameter space (colored region) of the lepto-phobic doublet (LPD) $G221$ model and the lepto-phobic triplet (LPT) $G221$ model, respectively, at $95\%$ CL  in the $\cos{\phi} - M_{W^\prime}$ and $M_{Z^\prime}- M_{W^\prime}$ planes after including EWPT constraints.

For both the LPD and LPT models, the allowed region in the $\cos{\phi} - M_{W^\prime}$ plane shows that direct search constraints favor small $\cos{\phi}$, which is expected because the $W^\prime$ coupling is proportional to $1/\sin{\phi}$, leading to small $W^\prime$ production rate in these regions. However, $\cos{\phi}$ can not be too small due to the perturbativity of the $g_2$ and $g_X$ coupling strength.  Conversely, in the $\cos{\phi} - M_{Z^\prime}$ plane,  small $\cos{\phi}$ is disfavored by direct LHC search constraints because $M_{Z^\prime} \simeq M_{W^\prime}/\cos{\phi}$. 

In the $M_{Z^\prime}- M_{W^\prime}$ plane of the Fig. \ref{parameter_constraints_LPD} and Fig. \ref{parameter_constraints_LPT}, we can see that the LPD model, with  $M_{W^\prime} \, \sim \, 2$ TeV, the EWPT constraints force $M_{Z^\prime} \, \geq \, 1.9$ TeV, while for the LPT model, with  $M_{W^\prime} \, \sim \, 2$ TeV, the EWPT constraints force $M_{Z^\prime} \, \geq \, 2.8$ TeV.

\section{Results and Predictions}
\label{sec:explain}

In this Section, we present our main results for explaining the $WZ$, $Wh$ and dijet excesses with our model. We discuss in turn the results for the $W'$ and the $Z'$ bosons.

\subsection{Results for $W^{\prime}$}

Before proceeding to the $W'$ predictions in our model, we note from Fig.~\ref{branchings_LPD} that there is appreciable branching of $W'$ into SM fermions. When resonantly produced in Drell-Yan processes, $W'\rightarrow l\nu$ and $Z'\rightarrow ll$ decays lead to tight constraints on the mass of $W'$~\cite{ATLAS:2014wra} and $Z'$~\cite{Aad:2014cka} bosons if their couplings to leptons resemble those between the SM $W,Z$ bosons to SM leptons. In our leptophobic scenario, the leptons are not charged under $SU(2)_R$ and the $W'\rightarrow l\nu$ decays are forbidden. Thus the current $W'$ mass constraint does not apply to our model. We will see later, however, the $Z'\rightarrow ll$ constraint is significant.


First let us focus on  the $WZ$ excess. The signal rate in the $WZ$ channel is evaluated as $\sigma_{W'}Br(W' \to WZ) A_{eff}$, and $A_{eff}$ is taken to be around 13\%, which is the diboson event selection efficiency~\cite{Aad:2015owa}. 
Including the event selection efficiency and luminosity, the signal cross section $\sigma_{W'}Br(W' \to WZ)$ should be around $3 \sim 15$ fb.
Theoretically, the $W'$ production cross-section $\sigma_{W'}$ in our G221 model can be obtained via the scaling from a NNLO `sequential SM' cross-section:
\be 
\sigma_{W'} = \sigma_{NNLO}\left(\frac{g_{W'L}}{g_{SM}}\right)^2,
\ee
where $g_{W'L}$ and $g_{SM}$ denote for the $W'$ and SM $W$ coupling to the quarks. We adopt the NNLO $W'$ production cross-section from Ref.~\cite{ATLAS:2014wra}, which is taken to be 292 fb for a `sequential SM' 2 TeV $W'$.

\begin{figure}
	\includegraphics[width=0.7\textwidth]{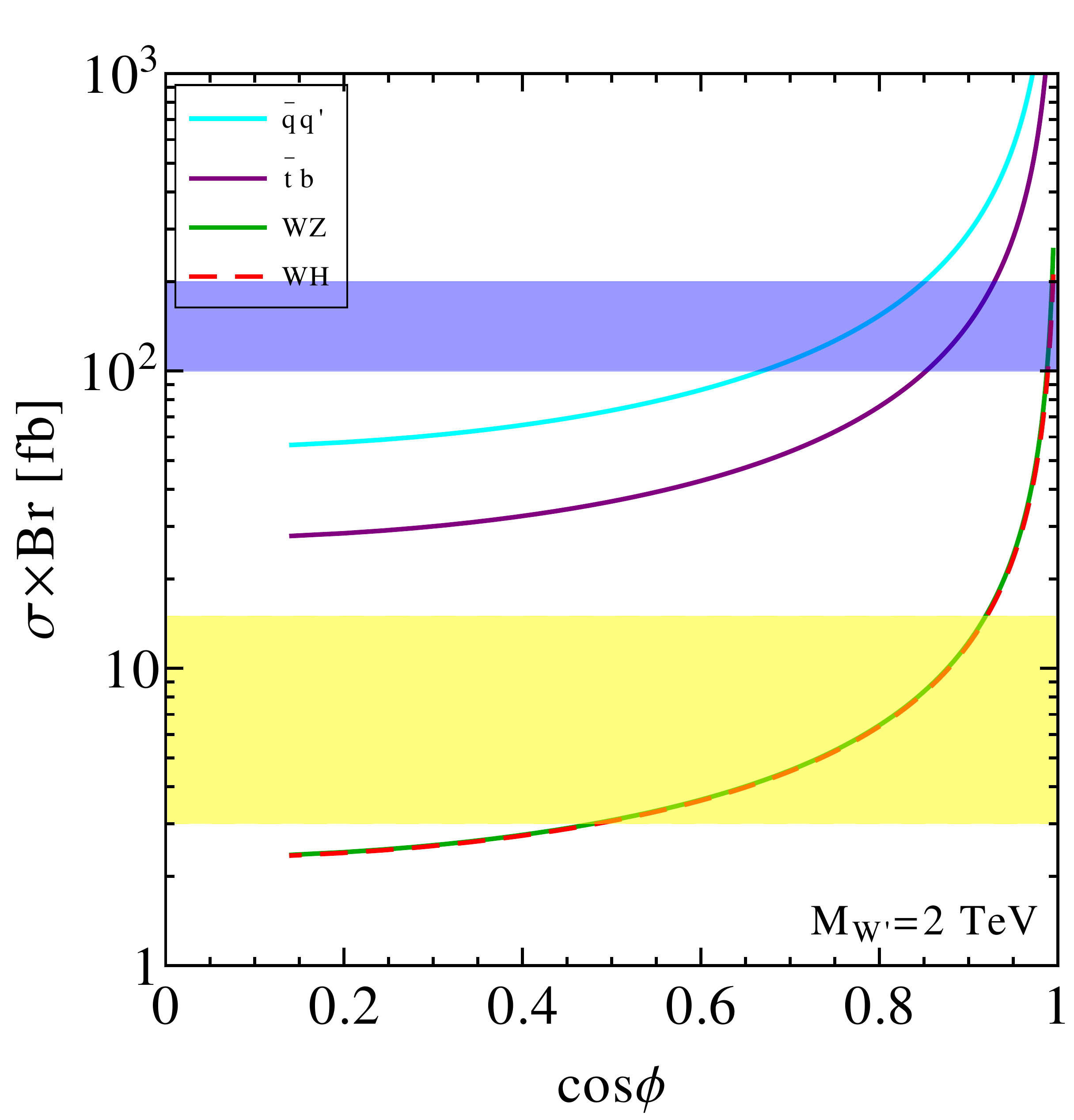}
	\caption{The cross section times branching to different channels for a $2$ TeV $W'$, as a function of $\cos{\phi}$. The coincident green and red lines denote the branching times cross section to $WZ$ and $Wh$ channels. The shaded yellow (blue) band denotes the region that is compatible with the ATLAS $WZ$ (dijet) excess. The $WZ$ and $Wh$ contours overlap due to the Goldstone equivalence theorem.}
\label{sigmaBrWprime}
\end{figure}	

Our results for $W^{\prime}$ are presented in Fig.~\ref{sigmaBrWprime}, where we show the cross section times branching for $W'$ in our model as a function of the mixing angle $\cos{\phi}$ for various channels. From top to bottom, the blue solid, purple solid, green solid, and red dashed lines show the model's prediction signal cross-section in the $q \overline{q'}$, $t \overline{b}$, $WZ$, and $Wh$ channels, respectively. The horizontal shaded yellow band denotes the parameter space compatible with the ATLAS $WZ$ excess with a cross section of $3 \sim 15$ fb. Thus a large range of the mixing angle value, $0.45< \cos{\phi} < 0.92 $, can explain the $WZ$ excess.

Given the $WZ$ signal, the equivalence theorem requires the  $W'\rightarrow Wh$ decay happen at a comparable rate to that of the longitudinal polarization of $Z$ in the $W'\rightarrow WZ$ process. Since $W'$ is heavy, the daughter $Z$ boson is boosted and dominated by its longitudinal mode. Hence BR($W'\rightarrow WZ$)$\approx$ BR($W'\rightarrow Wh$) and an equally large signal in the $Wh$ channel is predicted.

Interestingly, CMS has reported a $2\sigma$ up-fluctuation in the $e\nu b\bar{b}$ search~\cite{CMS:2015gla} that could arise from a 1.8-2.0 TeV $W'$ that decays into $Wh$. Since the 95\% confidence level uncertainty at $M_{W'}=2$ TeV is given~\cite{CMS:2015gla} at 8 fb, a 2$\sigma$ up-fluctuation approximately suggests an 8 fb $W'$ signal. Thus, approximately the same range of $\cos{\phi}$ that fits the $WZ$ excess would also fit this putative excess. We note that a similar excess in the  $\mu\nu b\bar{b}$ channel was not seen~\cite{CMS:2015gla} in the same analysis. More data will settle the question of whether this excess will be statistically established in the future.

As a benchmark point for these two channels, we choose $\cos{\phi} = 0.8$. At this point, the cross section times branching of the $W'$ boson to various channels are as follows:
\bea
\sigma(pp\rightarrow W') BR(W' \to \overline{q} q') \, = \, 150 \, {\rm fb}, \nonumber \\ 
\sigma(pp\rightarrow W') BR(W' \to \overline{t} b) \, = \, 71 \, {\rm fb}, \nonumber \\ 
\sigma(pp\rightarrow W') BR(W' \to WZ) \, = \, 6.3 \, {\rm fb}, \nonumber \\ 
\sigma(pp\rightarrow W') BR(W' \to Wh) \, = \, 6.3 \, {\rm fb}.
\label{eq:sigmaWBRs}
\eea

We now turn to the dijet channel. CMS also reported a $\sim$2$\sigma$ up-fluctuation~\cite{Khachatryan:2015sja} in quark-quark invariant mass at 1.8 TeV. By including the cut efficiency and luminosity, we obtain the dijet excess $\sigma(pp \to W' \to jj)$ around $100 \sim 200$ fb. If considered as an excess, it is consistent with a `sequential SM" $W'\rightarrow q q$ signal~\cite{Khachatryan:2015sja}. Our benchmark point yields 30\% of the $\sigma$BR($W'\rightarrow qq$) in comparison to the Sequential SM case, and fits in excess well. In Fig.~\ref{sigmaBrWprime}, the horizontal blue band shows the region with a dijet cross section around $100 \sim 200$ fb that explains the dijet excess.  Alternatively, even if the dijet data is interpreted as a bound that marginally excludes a Sequential SM $W'$ at 2 TeV, our 2 TeV $W'$ at the benchmark point can still be allowed due to its smaller couplings to the quarks. 

It is also interesting to note an associated single top $tb$ final state is also expected at 71 fb, as listed in Eq.~\ref{eq:sigmaWBRs}. While still below current LHC limits~\cite{Aad:2014xea}, it can be searched at future high statistics runs.

In conclusion, we see that after imposing the constraints from EWPT and current LHC data, we can explain the $WZ$, $Wh$ and dijet
excesses together for a range of values of the $SU(2)_R$ coupling strength $g_R$ in the range $0.47\sim 0.68$, 
which coresspondings to the range $0.66 < \cos\phi < 0.85$.

\subsection{Results for $Z^{\prime}$}

\begin{figure}
	\includegraphics[width=0.4\textwidth]{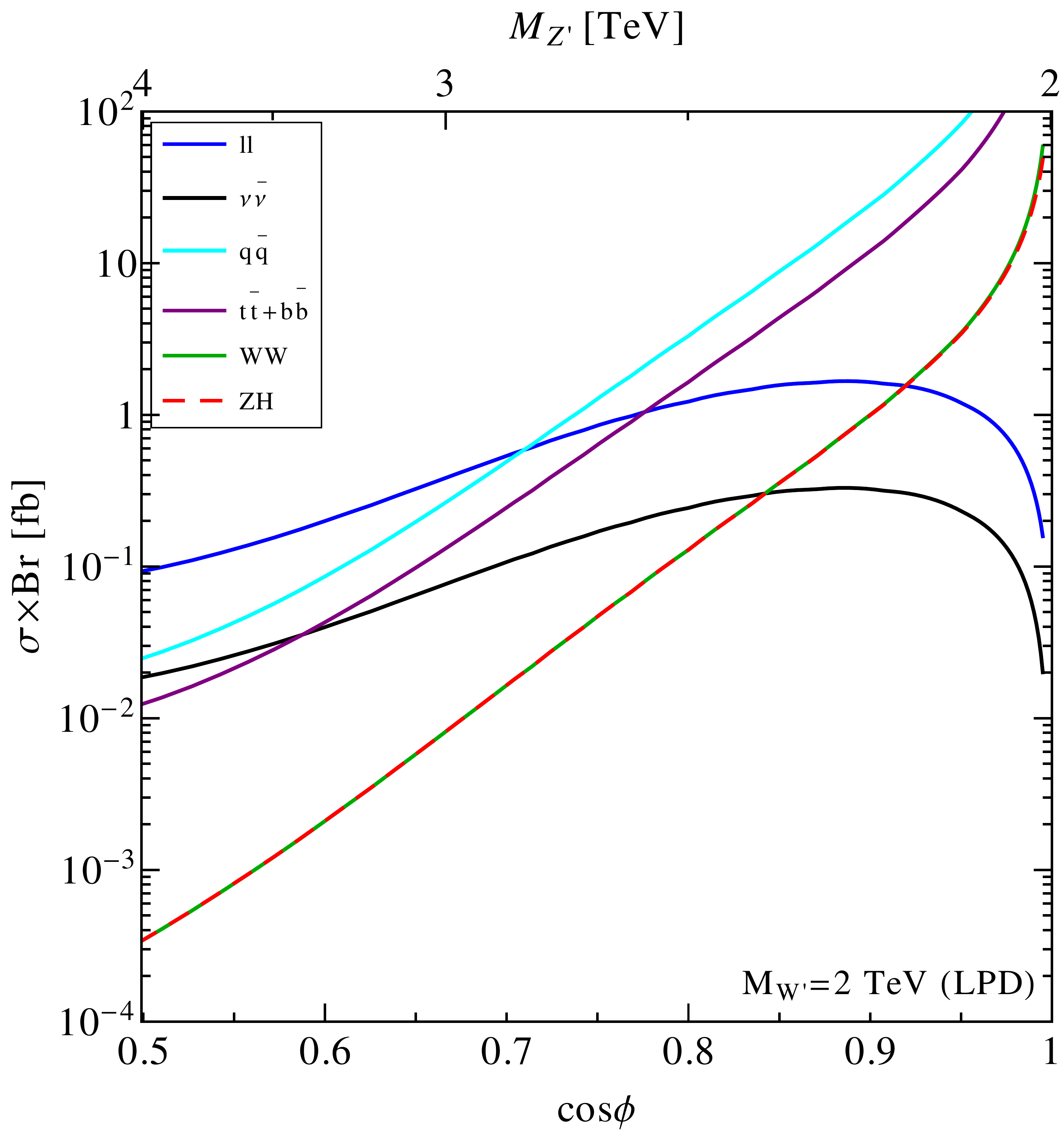}
	\includegraphics[width=0.4\textwidth]{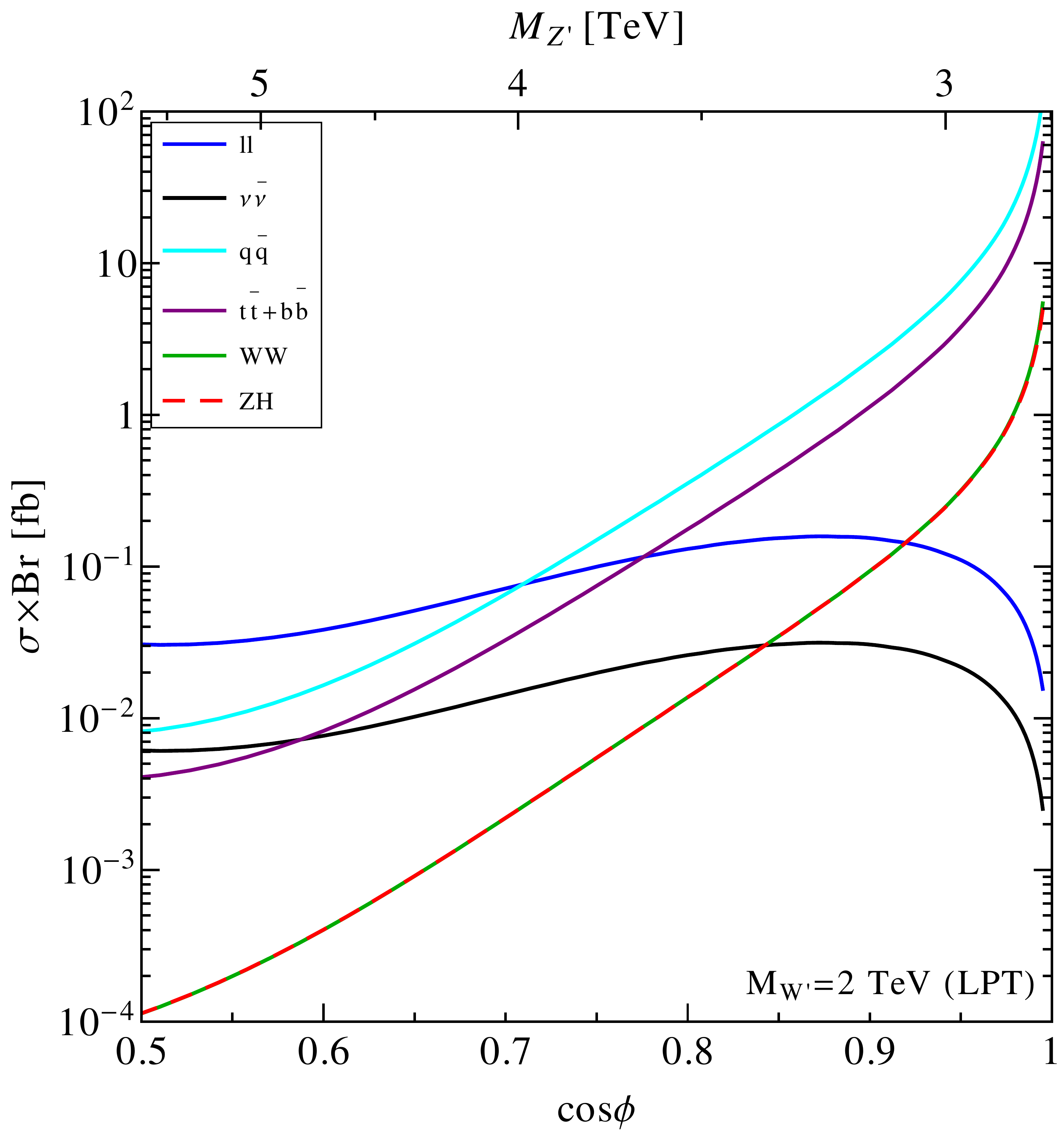}
	\caption{The cross section times branching to different channels for different mixing angle (or  $Z'$ mass) in the doublet model (left panel) and  in the triplet model (right panel), as a function of $\cos{\phi}$.}
\label{sigmaBrZprime}
\end{figure}

We now turn to constraints on the $Z^{\prime}$ boson in our model, and comment on the possibility of explaining the $WW$ excess. 
Since we know the favored region of the $W'$ mass and the mixing angle $\cos\phi$,  the favored $Z'$ mass and 
couplings could be fully predicted. 
Using our benchmark point with $\cos\phi = 0.7$, the $Z'$ mass is predicted to be 2.9 TeV for LPD model and 3.5 TeV for LPT model.
Our main results for the $Z^{\prime}$ boson in the benchmark point are summarized in Fig.~\ref{sigmaBrZprime}. In the left panel, we show the results for the LPD model as the function of the mixing angle, and the right panel shows the LPT model s the function of the mixing angle.

Firstly, we consider the dilepton constraint in the two charged lepton channel, relevant for  $Z'\rightarrow ll$. The leptons in our model are charged under $U(1)$ and thus $Z'\rightarrow ll$ processes can occur via the $Z'$ mixing with $Z$. In ATLAS's recent dilepton analysis~\cite{Aad:2014cka,Patra:2015bga}, the $Z'$ mass with `Sequential SM' couplings is constrained to 2.7$\sim$2.8 TeV at 95\% confidence level.

In the LPD case, the Figure \ref{zprime_mass} shows 
the  $Z^{\prime}$ mass is around $2 \sim 3 $ TeV in the favored $\cos{\phi}$ region from the $W^{\prime}$: $0.66 < \cos\phi < 0.85$ .
Fig.~\ref{sigmaBrZprime} shows the EWPT constraints allow a $Z'$ mass as low as 2.1 TeV for $M_{W'} \sim 2$ TeV. 
For this lower $Z'$ mass to be consistent with the dilepton search bound, $Z'$ must either have a small production cross section, i.e. smaller couplings to quarks, and/or a lower decay branching ratio into leptons than a sequential SM $Z'$ does. The combination of these two factors can be optimized by varying the $\cos{\phi}$ value.
At the benchmark point $\cos{\phi} = 0.7$, which corresponds to $M_{Z'} = 2.9$ TeV, the cross section times branching for the various channels are given below:
\bea
&&\sigma(pp\rightarrow Z') BR(Z' \to \overline{q} q') \, = \, 0.40 \, {\rm fb}, \nonumber \\ 
&&\sigma(pp\rightarrow Z') BR(Z' \to \overline{t} t(\overline{b} b)) \, = \, 0.20 \, {\rm fb}, \nonumber \\ 
&&\sigma(pp\rightarrow Z') BR(Z' \to ll) \, = \, 0.48 \, {\rm fb}, \nonumber \\ 
&&\sigma(pp\rightarrow Z') BR(Z' \to \nu \overline{\nu}) \, = \, 0.10 \, {\rm fb}, \nonumber \\ 
&&\sigma(pp\rightarrow Z') BR(Z' \to Zh) \, = \, 0.01 \, {\rm fb}, \nonumber \\ 
&&\sigma(pp\rightarrow Z') BR(Z' \to WW) \, = \, 0.01 \, {\rm fb}. 
\label{eq:sigmaBRs}
\eea
We found a minimal $Z'$ production at 21\% of the Sequential SM cross-section and a $Z'\rightarrow ll$ at 2.9 TeV that explains the recent dielectron event at the CMS~\cite{bib:cms2.9dielectron} and predict a similar resonance in di-muon channel. On the other hand, if we consider previous non-signal dilepton data, the LPD model has a lowest BR($Z'\rightarrow ll$)=$12\%$ within EWPT constraints, which are too large to evade $Z'\rightarrow ll$ constraints and a $Z'$ mass greater than 2.8 TeV is needed, which is possible with a smaller $\cos{\phi}$: $0.66 < \cos\phi < 0.72$.
From Figure \ref{sigmaBrWprime}, the smaller $\cos{\phi}$, the smaller the cross section times branching ratio for a 2 TeV $W'$. 
Therefore, to explain the the $WZ$, $Wh$ and dijet
excesses and escape the dilepton constraint in the LPD model, one needs to take the mixing angle to be around $0.66 < \cos\phi < 0.72$.

In the LPT model, the Figure \ref{zprime_mass} shows 
the  $Z^{\prime}$ mass is around $3 \sim 5 $ TeV in the favored $\cos{\phi}$ region.
This $Z^{\prime}$ is beyond the search limits of the current dilepton bound. 
As shown in Fig.~\ref{parameter_constraints_LPT}, the EWPT constraints in the LPT scenario allow $M_{Z'}\ge 2.8$ TeV for $M_{W'} \sim 2$ TeV.  
Therefore the LPT model is totally consistent with the ATLAS dilepton bound, and can also accommodate for the CMS $e^+e^-$ event at 2.9 TeV.
However, such a large $Z'$ mass would be unsuitable to explain the diboson  $WW$ excess.

In conclusion, we find that, the LPD scenario predicts a $Z'$ boson with mass around $2 \sim 3$ TeV, which is compatible with EWPT but is tightly constrained by the dilepton searches, except the parameter region with $0.66 < \cos\phi < 0.72$, which corresponds to $M_{Z'} > 2.8 $ TeV.
We also note that the LPT scenario predicts a $Z'$ boson with mass around $3 \sim 5$ TeV that is completely compatible with EWPT and LHC dilepton constraints, which, however, would be irrelevant for the recent $WW$ diboson excess.

\section{Conclusions}
\label{sec:conclusion}

We investigated the prospects of the leptophobic $SU(2)_L \times SU(2)_R \times U(1)_{B-L}$ model as a potential explanation to the diboson and $Wh$ excesses. 
In our discussion, we fixed the $W'$ mass to be 2 TeV.
Within the electroweak precision data limits, we found that 
to explain the $WZ$, $Wh$ and dijet excesses together, the $SU(2)_R$ coupling strength $g_R$ favors the range of $0.47 \sim 0.68$ and a range for mixing angle $0.66<\cos\phi<0.85$. 
We noticed that the $Z'$ mass and couplings are determined by the two parameters appeared in the $W'$ sector. 
Therefore, given the favored region to explain the excesses, the $Z'$ masses are determined to be around $2 \sim 3$ TeV for LPD and $3 \sim 5$ TeV for LPT model, and the $Z'$ decay widths to the dilepton, dijet, and gauge bosons are predicted.  
We found the ATLAS $WW$ and $ZZ$ excesses are unlikely to arise from the heavy $Z'$ from this model due to a much heavier $Z'$ mass in the LPT model. Within electroweak precision limits, the benchmark LPD point can explain the CMS's recent dielectron event at 2.9 TeV, while the heavier LPT $Z'$ could also be consistent with 2.9 TeV mass, or evade dilepton bounds even in case if future data do not establish the 2.9 TeV excess.

As a model independent check, the leptonic decay of the $W' \to WZ$ bosons would lead to a $3l+\met$ final state with the same invariant mass around 2 TeV. No significant excess has been reported in this channel, and the current CMS~\cite{Khachatryan:2014xja} data place a constraint of $\sigma\times \text{BR}(W'\rightarrow 3l\nu)$ below 0.1 fb for $M_{W'}=2$ TeV. Given the SM $WZ$ leptonic decay branching fractions, the relative size to the four jet final state is 0.03. If the four jet $WZ$ excess persists, an associated $\sigma_{W'}\text{BR}(W'\rightarrow 3l\nu)$ excess at 0.2 fb is expected. Also, no significant deviation from the SM was observed from ATLAS's recent analysis~\cite{Aad:2015ufa} of the semileptonic $WZ/WW\rightarrow l\nu jj$ channel. It is noted that many of the aforementioned up-fluctuations are statistically limited in the current data and LHC run 2 updates will greatly help confirm or clarify the excesses.

In summary, the recent tantalizing excesses in the $WZ$, $Wh$, and dijet channels can be accommodated with the LPT model and a limit range of parameter space of the LPD model, in a manner consistent with both EWPT and LHC constraints.

\vspace{0.5in}
{\bf{Note after published}}: With a 2.9 TeV $Z'$  event observed in Ref.~\cite{bib:cms2.9dielectron}, we updated our paper by discussing the possible 2.9 TeV $Z'$ signature together with the 2 TeV $W'$ excess.

\section*{Acknowledgements}

We would like to thank the CETUP 2015 Dark Matter Workshop in South Dakota for providing a stimulating atmosphere where this work was conceived and concluded. Y.G. thanks the Mitchell Institute for Fundamental Physics and Astronomy for support. T.G. is supported by DOE Grant DE-FG02-13ER42020.  K.S. is supported by NASA Astrophysics Theory Grant NNH12ZDA001N. The research of JHY is supported by the National Science Foundation under Grant Numbers PHY-1315983 and PHY-1316033.

\appendix

\section{Heavy Gauge Boson Decay Width}

The partial decay width of $V' \to \bar{f}_{1}f_{2}$ is
\be
\Gamma_{V'\to\bar{f_{1}}f_{2}}=\frac{M_{V'}}{24\pi}\beta_{0}\left[(g_{L}^{2}+g_{R}^{2})\beta_{1}+6g_{L}g_{R}\frac{m_{f_{1}}m_{f_{2}}}{M_{V'}^2}\right]\Theta(M_{V'}-m_{f_{1}}-m_{f_{2}})\,,
\label{eq:v_width}
\ee
where 
\bea
\beta_{0}&=&\sqrt{1-2\frac{m_{f_{1}}^{2}+m_{f_{2}}^{2}}{M_{V'}^{2}}+\frac{(m_{f_{1}}^{2}-m_{f_{2}}^{2})^{2}}{M_{V'}^{4}}}, \nonumber \\
\beta_{1}&=&1-\frac{m_{f_{1}}^{2}+m_{f_{2}}^{2}}{2M_{V'}^{2}}-\frac{(m_{f_{1}}^{2}-m_{f_{2}}^{2})^{2}}{2 M_{V'}^{4}}.
\eea
The color factor $N_c$ is not included and
the top quark decay channel only open when the $Z^\prime$ and $W^\prime$ masses are heavy. 

The partial decay width of $V'\to V_{1}V_{2}$ is
\be
\Gamma_{V'\to V_{1}V_{2}}=\frac{M_{V'}^{5}}{192\pi M_{V_{1}}^{2}M_{V_{2}}^{2}}g_{V'V_{1}V_{2}}^{2}\beta_{0}^{3}\beta_{1}\Theta(M_{V'}-M_{V_{1}}-M_{V_{2}})\,,
\ee
where 
\bea
\beta_{0}&=&\sqrt{1-2\frac{M_{V_{1}}^{2}+M_{V_{2}}^{2}}{M_{V'}^{2}}+\frac{(M_{V_{1}}^{2}-M_{V_{2}}^{2})^{2}}{M_{V'}^{4}}},\nonumber \\
\beta_{1}&=&1+10\frac{M_{V1}^{2}+M_{V2}^{2}}{2M_{V'}^{2}}+\frac{M_{V_{1}}^{4}+10M_{V1}^{2}M_{V_{2}}^{2}+M_{V_{2}}^{4}}{M_{V'}^{4}}.
\eea
The partial decay width of $V'\to V_{1}H$ (where $V_1=W$ or $Z$ boson and $H$ is the lightest Higgs boson) is
\be
\Gamma_{V'\to V_{1}H}=\frac{M_{V'}}{192\pi}\frac{g_{V'V_{1}H}^{2}}{M_{V_{1}}^{2}}\beta_{0}\beta_{1}\Theta(M_{V'}-M_{V_{1}}-M_{V_{2}})\,,
\ee
where 
\bea
\beta_{0}&=&\sqrt{1-2\frac{M_{V_{1}}^{2}+m_{H}^{2}}{M_{V'}^{2}}+\frac{(M_{V_{1}}^{2}-m_{H}^{2})^{2}}{M_{V'}^{4}}}, \nonumber \\
\beta_{1}&=&1+\frac{10M_{V_{1}}^{2}-2m_{H}^{2}}{2M_{V'}^{2}}+\frac{(M_{V_{1}}^{2}-m_{H}^{2})^{2}}{M_{V'}^{4}}.
\eea


\end{document}